\documentclass[graybox]{svmult}
\usepackage{mathptmx}       % selects Times Roman as basic font
\usepackage{helvet}         % selects Helvetica as sans-serif font
\usepackage{courier}        % selects Courier as typewriter font
\usepackage{makeidx}         % allows index generation
\usepackage{graphicx}        % standard LaTeX graphics tool
\usepackage{multicol}        % used for the two-column index
\usepackage[bottom]{footmisc}% places footnotes at page bottom
\usepackage{amsmath,amssymb,amsfonts,bbold}        % AMS math packages
\makeindex             % used for the subject index
\newcommand{\dder}[2][]{%
	\ifthenelse{\equal{#1}{}}{%
		\frac{\mathrm{d}}{\mathrm{d} #2}%
	}{%
		\frac{\mathrm{d} #1}{\mathrm{d} #2}%
	}%
}

\newcommand{\inner}[2]{\langle#1,#2\rangle}

\begin{document}
\title*{On Super Non-Abelian T-Duality of Symmetric and Semi-Symmetric Coset Sigma Models}
% Use \titlerunning{Short Title} for an abbreviated version of
% your contribution title if the original one is too long
\author{Daniele Bielli}
% Use \authorrunning{Short Title} for an abbreviated version of
% your contribution title if the original one is too long
\institute{Daniele Bielli \at University of Surrey (UK) and Milano-Bicocca (Italy), \email{d.bielli@surrey.ac.uk}
}
%
% Use the package "url.sty" to avoid
% problems with special characters
% used in your e-mail or web address
%
\maketitle

\abstract{We review the algebraic approach to super non-Abelian T-Duality considered in \cite{snatd}, focusing on symmetric and semi-symmetric coset spaces on $G/H$. We discuss a potential impediment, appearing in these models when integrating out the gauge field in favour of the dual variables. This process cannot be performed in general and we isolate the obstruction, highlighting three cases in which a solution can be found. After writing the T-dual action we provide solution for two specific models. The first based on the symmetric space $S^{3} \simeq SO(4)/SO(3)$, well-known in the literature, the second on the semi-symmetric coset $OSp(1|2)/SO(1,1)$, a Green-Schwarz-like string sigma model satisfying the supergravity torsion constraints. \footnote{DMUS-MP-22/18. Contribution to the proceedings of the MATRIX Research Program \textit{2D Supersymmetric Theories and Related Topics}.}
}

\section{Introduction}

\label{sec:1}
We start by fixing notation on semi-symmetric space sigma models, as relevant formulae can be reduced to the ones for symmetric spaces. We consider two-dimensional sigma models involving a generic Lie (super)group $G$ with associated Lie (super)algebra $\mathfrak{g}$ and defined by considering smooth maps $g\in C^\infty(\Sigma,G)$ from a two-dimensional Lorentzian world-sheet $\Sigma$ to $G$. The main building block of these models is the pull-back to $\Sigma$, via $g$, of the Maurer-Cartan form $j := g^{-1} \mathrm{d} g \in \Omega^{1}(\Sigma,\mathfrak{g})$, which is by construction invariant under global $G_{L}$ transformations $g\rightarrow g_{0}^{-1}g$ and satisfies the Maurer-Cartan flatness condition $F_{j}:=\mathrm{d}j + \frac{1}{2}[j,j]=0$, with $\mathrm{d}$ and $\Omega^p(\Sigma,\mathfrak{g})$ respectively denoting exterior derivative and $\mathfrak{g}$-valued $p$-forms on $\Sigma$. Semi-symmetric space sigma models are then coset models on $G/H$ for which $H$, subgroup of $G$ with associated Lie algebra $\mathfrak{h}$, arises as the fixed point set of an automorphism of $G$ of order four. At the Lie algebra level this implies the existence of an automorphism $\sigma: \mathfrak{g} \rightarrow \mathfrak{g}$ whose fourth power is the identity. Using the latter, one can define projectors from $\mathfrak{g}$ to four subspaces $\mathfrak{h}:= P_{\mathfrak{h}}(\mathfrak{g})$, \,  $\mathfrak{p}:= P_{\mathfrak{p}}(\mathfrak{g})$,  \,  $\mathfrak{m}:= P_{\mathfrak{m}}(\mathfrak{g})$,  \, $\mathfrak{q}:= P_{\mathfrak{q}}(\mathfrak{g})$ and find an orthogonal decomposition $\mathfrak{g}\simeq \mathfrak{h} \oplus \mathfrak{p}\oplus \mathfrak{m} \oplus \mathfrak{q}$. The subspaces $\mathfrak{h},\mathfrak{m}$ and $\mathfrak{p},\mathfrak{q}$ are respectively purely bosonic and purely fermionic.
In turn, the current $j$ decomposes as $j=A+p+m+q$, with the components transforming as $A\rightarrow h^{-1}Ah + h^{-1}\mathrm{d}h$ and $\{ p,m,q \} \rightarrow h^{-1} \{ p,m,q \} h$ under the local $H_{R}$ action $g\rightarrow gh$. A locally $H_{R}$-invariant and globally $G_{L}$-invariant action is then constructed out of $p,m,q$ as $S_{SS}=\tfrac{1}{2}\int_{\Sigma}\langle m,\star m\rangle+\tfrac{1}{2}\int_{\Sigma}\langle p,q\rangle$. Wedge product is understood and $\star$ is the Hodge star operator on $\Sigma$ with respect to the world-sheet metric, satisfying $\star^2=1$ on $\Omega^{1}(\Sigma,\mathfrak{g})$. Brackets denote an Ad-invariant, non-degenerate graded-symmetric bilinear form on $\mathfrak{g}$ compatible with the decomposition.

T-duality can then be performed, assuming a topologically trivial world-sheet, by gauging a sub-(super)group $K_{L} \subseteq G_{L}$ of the isometries with a gauge field $\omega \in \Omega^{1}(\Sigma,\mathfrak{k}_{L})$ and introducing an extra term enforcing the flatness $F_{\omega}:=\mathrm{d}\omega+\tfrac{1}{2}[\omega,\omega]=0$ by means of Lagrange multipliers $\Lambda \in C^{\infty}(\Sigma,\mathfrak{k}_{L})$, which transform as $\Lambda \rightarrow h^{-1}\Lambda h$ under local $H_{R}$. Integrating out the multipliers, $\omega$ is set to be pure gauge and the initial model can be recovered, while integrating out $\omega$ a T-dual model is obtained in which the multipliers play the role of dual coordinates. The gauged action, with Lagrange multipliers, reads $S_{SS}^{\omega}\ =\ \tfrac12\int_\Sigma\inner{m_\omega}{\star m_\omega}+\tfrac12\int_\Sigma\inner{p_\omega}{q_\omega}+\tfrac12\int_\Sigma\inner{D(j_\omega)}{j_\omega}+\int_\Sigma\inner{\tilde\Lambda}{F_{j_\omega}}$. Where $\tilde{\Lambda}:=g^{-1}\Lambda g+g^{-1}D(g)$, while $j_{\omega}:= j +g^{-1}\omega g$ and $A_{\omega},p_{\omega}, m_{\omega}, q_{\omega}$ are its projections on the subspaces. $D:\mathfrak{g}\rightarrow \mathfrak{g}$ defines a deformation of the initial model introduced in \cite{borsato2016,borsato2017,borsato2018}, which will be set to zero in our examples. We shall dualise the whole isometry groups and also choose gauge  $g=\mathbb{1}$. The EOM for $\omega$ reads ${\star m_\omega}-\tfrac12p_\omega+\tfrac12q_\omega+\nabla_{j_\omega}\tilde\Lambda-D(j_\omega)\ =\ 0~$
and its projections on $\mathfrak{p},\mathfrak{m},\mathfrak{q}$ can be solved for $p_{\omega}, m_{\omega}, q_{\omega}$ \cite{snatd}. The projection on $\mathfrak{h}$ reads
\begin{equation}\label{h-projection-EOM}
[\tilde\Lambda_\mathfrak{q},p_\omega]
+D_{\tilde\Lambda_\mathfrak{m}}(m_\omega)+[\tilde\Lambda_\mathfrak{p},q_\omega]\ =\ \nabla_{A_{\omega}}\tilde\Lambda_\mathfrak{h}
\end{equation}
with $D_{\tilde{\Lambda}_{\mathfrak{m}}}:=(D+ad_{\tilde{\Lambda}_{\mathfrak{m}}})(m_{\omega})$ and $\nabla_{j_{\omega}}:=\mathrm{d}+ad_{j_{\omega}}$, and cannot be generally solved for $A_{\omega}$ due to the absence of linear terms. This issue is a direct consequence of local $H_{R}$-invariance for physically relevant initial actions and forces a case-by-case study.

\section{Solving the EOM - two simple examples}
Exploiting the solution for $p_{\omega},m_{\omega},q_{\omega}$, equation \eqref{h-projection-EOM} can be rearranged as
\begin{equation}\label{EOM-A-WZ}
W(A_{\omega})+Z(\star A_{\omega}) = \zeta
\end{equation}
where $W,Z: \mathfrak{h}\rightarrow \mathfrak{h}$ and $\zeta\in \Omega^{1}(\Sigma,\mathfrak{h})$ have explicit expressions
\begin{equation}
\begin{cases}
W :=ad_{\tilde{\Lambda}_{\mathfrak{h}}}+N+(D_{\tilde{\Lambda}_{\mathfrak{m}}}-M^{\dagger})\circ \sum_{k=0}^{\infty}S^{2k+1}\circ (D_{\tilde{\Lambda}_{\mathfrak{m}}}+M)
\\
Z:=(D_{\tilde{\Lambda}_{\mathfrak{m}}}-M^{\dagger})\circ \sum_{k=0}^{\infty}S^{2k}\circ (D_{\tilde{\Lambda}_{\mathfrak{m}}}+M)
\\
\zeta:=\mathrm{d}\tilde{\Lambda}_{\mathfrak{h}}+\xi+(D_{\tilde{\Lambda}_{\mathfrak{m}}}-M^{\dagger})\circ(\star+S)\circ \sum_{k=0}^{\infty}S^{2k}\circ (\mathrm{d}\tilde{\Lambda}_{\mathfrak{m}}+\chi)
\end{cases}
\end{equation}
and we further defined $S:= ad_{\tilde{\Lambda}_{\mathfrak{h}}}+L$, with $L:=ad_{\tilde{\Lambda}_{\mathfrak{p}}}\circ \mathcal{O}_{1}+ad_{\tilde{\Lambda}_{\mathfrak{q}}}\circ \mathcal{O}_{2}$ and
\begin{equation}
\begin{cases}
M:=ad_{\tilde{\Lambda}_{\mathfrak{p}}}\circ \mathcal{O}_{3}+ad_{\tilde{\Lambda}_{\mathfrak{q}}}\circ \mathcal{O}_{4}
\\
N:=ad_{\tilde{\Lambda}_{\mathfrak{q}}}\circ \mathcal{O}_{3}+ad_{\tilde{\Lambda}_{\mathfrak{p}}}\circ \mathcal{O}_{4}
\\
\xi:=\mathcal{O}_{3}^{\dagger}(\mathrm{d}\tilde{\Lambda}_{\mathfrak{q}})+\mathcal{O}_{4}^{\dagger}(\mathrm{d}\tilde{\Lambda}_{\mathfrak{p}})
\\
\chi:=\mathcal{O}_{1}^{\dagger}(\mathrm{d}\tilde{\Lambda}_{\mathfrak{q}})+\mathcal{O}_{2}^{\dagger}(\mathrm{d}\tilde{\Lambda}_{\mathfrak{p}})
\end{cases}
\qquad
\begin{cases}
\mathcal{O}_{1}:= R_{11}\circ ad_{\Lambda_{\mathfrak{q}}}+R_{12}\circ ad_{\Lambda_{\mathfrak{p}}}
\\
\mathcal{O}_{2}:= R_{21}\circ ad_{\Lambda_{\mathfrak{q}}}+R_{22}\circ ad_{\Lambda_{\mathfrak{p}}}
\\
\mathcal{O}_{3}:= R_{12}\circ ad_{\Lambda_{\mathfrak{q}}}+R_{11}\circ ad_{\Lambda_{\mathfrak{p}}}
\\
\mathcal{O}_{4}:= R_{22}\circ ad_{\Lambda_{\mathfrak{q}}}+R_{21}\circ ad_{\Lambda_{\mathfrak{p}}}
\end{cases}
\end{equation}
We refer to \cite{snatd} for the expressions of $R_{ij}$ and clarify that any $\mathcal{O}^{\dagger}$ is defined, with respect to the inner product $\langle \mathcal{O}^{\dagger}(X),Y \rangle = \langle X, \mathcal{O}(Y) \rangle$ for any two 1-forms $X,Y$, using $\langle R_{12}(X),Y \rangle = -\langle X,R_{12}(Y) \rangle$, $\langle R_{21}(X),Y \rangle = -\langle X,R_{21}(Y) \rangle$, $\langle R_{11}(X),Y \rangle = -\langle X,R_{22}(Y) \rangle $. As anticipated, all the above formulae reduce to those for symmetric spaces, in which $\mathfrak{g}\simeq \mathfrak{h}\oplus \mathfrak{m}$, by simply setting to zero any element in $\mathfrak{p}$ and $\mathfrak{q}$. In symmetric supercosets, $\mathfrak{h}$ and $\mathfrak{m}$ contain both bosonic and fermionic generators.

The possibility of solving \eqref{EOM-A-WZ} for $A_{\omega}$, depends on the invertibility of $W$ and $Z$ and in turn on the structure of the algebra. This potential obstruction has been mentioned in \cite{lozano,borsato2017,borsato2018} and problems in the presence of fermions have been discussed in \cite{grassi}. We now highlight three situations in which the equation can be solved: in the first two cases $W$ and $1\pm ZW^{-1}$ or $Z$ and $1\pm WZ^{-1}$ are invertible, while in the third one neither $W$ nor $Z$ is invertible, but their sum and difference are
\begin{equation}\label{Solution-EOM}
A_{\omega}=\tfrac{1}{2}(\zeta+\star \zeta)B_{+}+\tfrac{1}{2}(\zeta-\star \zeta)B_{-} \quad \mbox{with} \, 
\begin{cases}
B_{\pm}:=W^{-1}[(1\pm ZW^{-1})^{-1}]
\\
B_{\pm}:=\pm Z^{-1}[(1\pm WZ^{-1})^{-1}]
\\
B_{\pm}:=(W\pm Z)^{-1}
\end{cases}
\end{equation}
Writing $A_{\omega}= \star \alpha + \beta$ with $\alpha:=\tfrac{1}{2}(\rho_{+}B_{+}-\rho_{-}B_{-})$, $\beta:=\tfrac{1}{2}(\rho_{+}B_{+}+\rho_{-}B_{-})$ and $\rho_{\pm}:=\mathrm{d}\tilde{\Lambda}_{\mathfrak{h}}+\xi\pm(D_{\tilde{\Lambda}_{\mathfrak{m}}}-M^{\dagger})\circ \sum_{k=0}^{\infty}(\pm S)^{k}\circ (\mathrm{d}\tilde{\Lambda}_{\mathfrak{m}}+\chi)$, one finally obtains the full T-dual action $\tilde{S}_{SS}=\int_{\Sigma} \tilde{g} + \tilde{B}$ from the hybrid one in \cite{snatd}
\begin{align}
\tilde{g}&:=\tfrac{1}{2}  \langle \lambda_{-}, \frac{1}{1-S} \star\lambda_{+} \rangle - \langle \nabla_{\beta}\tilde{\Lambda}_{\mathfrak{h}}+\mathcal{O}_{3}^{\dagger}(\nabla_{\beta}\tilde{\Lambda}_{\mathfrak{q}})+\mathcal{O}_{4}^{\dagger}(\nabla_{\beta}\tilde{\Lambda}_{\mathfrak{p}}),\star \alpha \rangle
\\
\tilde{B}&:=\tfrac{1}{2} \langle \lambda_{-},\frac{1}{1-S} \lambda_{+} \rangle + \langle \tilde{\Lambda}_{\mathfrak{h}} , F_{\beta}-\tfrac{1}{2}[\alpha,\alpha] \rangle +\tfrac{1}{2} \langle \alpha, N(\alpha)\rangle+
\notag\\
& \, \, + \tfrac{1}{2} \langle \nabla_{\beta}\tilde{\Lambda}_{\mathfrak{p}}, R_{21}(\nabla_{\beta}\tilde{\Lambda})_{\mathfrak{p}} +R_{22}(\nabla_{\beta}\tilde{\Lambda})_{\mathfrak{q}}\rangle + \tfrac{1}{2} \langle \nabla_{\beta}\tilde{\Lambda}_{\mathfrak{q}}, R_{11}(\nabla_{\beta}\tilde{\Lambda})_{\mathfrak{p}} +R_{12}(\nabla_{\beta}\tilde{\Lambda})_{\mathfrak{q}}\rangle
\notag
\end{align}
with $\lambda_{\pm}:=\nabla_{\beta\pm\alpha}\tilde{\Lambda}_{\mathfrak{m}}+\mathcal{O}_{1}^{\dagger}(\nabla_{\beta\pm\alpha}\tilde{\Lambda}_{\mathfrak{q}})+\mathcal{O}_{2}^{\dagger}(\nabla_{\beta\pm\alpha}\tilde{\Lambda}_{\mathfrak{p}})-D(\beta\pm\alpha)$.

We shall now solve \eqref{EOM-A-WZ} in two examples. The first one involves dualisation of the $SO(4)$ isometry of the coset space $SO(4)/SO(3)\simeq S^{3}$. This model has been studied in the literature \cite{delaossa, lozano} and it is thus interesting to understand how the procedure goes through from the purely algebraic point of view. Given the $SO(4)$ algebra
\begin{equation}
[R_{IJ},R_{KL}]=-\tfrac{i}{2}(\delta_{IK}R_{JL}-\delta_{JK}R_{IL}-\delta_{IL}R_{JK}+\delta_{JL}R_{IK})
\end{equation}
one can separate the $\mathfrak{h}=SO(3)=\{ H_{i}:=-\tfrac{1}{2}\varepsilon_{i}{}^{jk}R_{jk} \}$ subalgebra from the rest $\mathfrak{m}=\{ M_{i}:= R_{i4} \}$ by using indices $I,J = \{1,2,3,4 \}$ and $i,j = \{ 1,2,3 \}$. Indices are raised and lowered using the Euclidean metric $\delta_{IJ}$. Generators in $\mathfrak{h}$ and $\mathfrak{m}$ satisfy
\begin{equation}
[H_{i},H_{j}]=\tfrac{i}{2}\varepsilon_{ij}{}^{k}H_{k} \qquad [M_{i},H_{j}]=\tfrac{i}{2}\varepsilon_{ij}{}^{k}M_{k} \qquad [M_{i},M_{j}]=\tfrac{i}{2}\varepsilon_{ij}{}^{k}H_{k}
\end{equation}
Writing the multipliers, i.e. the dual coordinate, as $\Lambda_{\mathfrak{h}}:=\tilde{y}^{i}H_{i}$ and $\Lambda_{\mathfrak{m}}:=\tilde{x}^{i}M_{i}$, recalling that $A_{\omega}:=A_{\omega}^{i}H_{i}$ and using the above commutators, one finds
\begin{equation}
W(A_{\omega})=v^{i}\varepsilon_{ij}{}^{k}A_{\omega}^{j}H_{k} \qquad \mbox{with} \qquad v^{i}:=\biggl[ \tfrac{i}{2}\tilde{y}^{i}+\frac{i(\tilde{x} \cdot \tilde{y})}{2(4-\tilde{y}^2)}\tilde{x}^{i} \biggr]
\end{equation}
\begin{align}
Z(\star A_{\omega})&=\tfrac{1}{4(\tilde{y}^2-4)}\bigl\{ [(\tilde{y}^2-4)\tilde{x}^2-(\tilde{y}\cdot\tilde{x})^2]\delta_{j}^{k}+
\\
& \qquad \qquad +[(\tilde{y}\cdot \tilde{x})\tilde{y}_{j}-(\tilde{y}^2-4)\tilde{x}_{j}] \tilde{x}^{k}+[(\tilde{y}\cdot\tilde{x})\tilde{x}_{j}-\tilde{x}^2\tilde{y}_{j}]\tilde{y}^{k}\bigr\}\star A_{\omega}^{j}H_{k}
\notag
\end{align}
It is clear that $v_{k}$ and $\tilde{x}_{k}$ respectively lie in the kernels of $W_{j}^{k}$ and $Z_{j}^{k}$, but it is also not hard to find that $W\pm Z$ can be inverted, so that $B_{\pm}:=(W\pm Z)^{-1}$ reads
\begin{align}
(B_{\pm})_{k}{}^{l}&=a_{1}^{\pm}\delta_{k}^{l}+\tilde{x}_{k}(a_{2}^{\pm}\tilde{x}^{l}+a_{3}^{\pm}\tilde{y}^{l})+\tilde{y}_{k}(a_{4}^{\pm}\tilde{x}^{l}+a_{5}^{\pm}\tilde{y}^{l})+\varepsilon_{ak}{}^{l}(a_{6}^{\pm}\tilde{x}^{a}+a_{7}^{\pm}\tilde{y}^{a})+
\\
& +\tilde{x}^{a}\tilde{y}^{b}\varepsilon_{ab}{}^{l}(a_{8}^{\pm}\tilde{x}_{k}+a_{9}^{\pm}\tilde{y}_{k})+\tilde{x}^{a}\tilde{y}^{b}\varepsilon_{abk}(a_{10}^{\pm}\tilde{x}^{l}+a_{11}^{\pm}\tilde{y}^{l})+ a_{12}^{\pm}\tilde{x}^{a}\tilde{y}^{b}\varepsilon_{abk} \tilde{x}^{c}\tilde{y}^{d}\varepsilon_{cd}{}^{l}
\notag
\end{align}
with $a_{1}^{\pm},...,a_{12}^{\pm}$ complicated functions of $\tilde{x}^2,\tilde{y}^2,(\tilde{x}\cdot \tilde{y})$ which we omit for brevity. 

The second example has to do with super T-dualisation of the $OSp(1|2)$ isometry of the semi-symmetric space $OSp(1|2)/SO(1,1)$. The interest in such a coset, which has also been considered in the context of holography \cite{verlinde}, is due to its structure, which is that of a 2d Green-Schwarz string sigma model satisfying the torsion constraints of supergravity. For this reason, dualising such a model would not only represent a natural next step to the super T-dualisation of the Principal Chiral Model on $OSp(1|2)$, for which in \cite{snatd} it was argued that T-duality breaks the supergravity constraint, but also a concrete example of super T-duality on supercosets, that could be compared to \cite{borsato2016,borsato2017,borsato2018}.
Given the $OSp(1|2)$ algebra in light-cone notation
\begin{align}
\{ Q_{+},Q_{+} \}=L_{++} \qquad \{ Q_{-}&,Q_{-} \}=L_{--} \qquad \{ Q_{+},Q_{-} \}=L_{+-}
\notag \\
[L_{+-},L_{\pm\pm}]=\pm iL_{\pm\pm} \quad & \quad  [L_{++},L_{--}]=-2i L_{+-}
\\ \notag 
[L_{++},Q_{-}]=-iQ_{+} \qquad [L_{--}&,Q_{+}]=iQ_{-} \qquad [L_{+-},Q_{\pm}]=\pm \tfrac{i}{2}Q_{\pm}
\end{align}
the four subspaces are $\quad \mathfrak{h}=\{L_{+-}\} \quad \mathfrak{p}=\{ Q_{+} \} \quad \mathfrak{m}=\{ L_{++},L_{--} \} \quad \mathfrak{q}=\{ Q_{-} \}$. Then, writing the Lagrange multipliers, i.e. the dual coordinates, as $\Lambda_{\mathfrak{h}}:=\tilde{y} \, L_{+-}$, \, $\Lambda_{\mathfrak{p}}:=\tilde{\theta}^{+}Q_{+}$, \,  $\Lambda_{\mathfrak{m}}:=\tilde{x}^{++}L_{++}+\tilde{x}^{--}L_{--}$, \, $\Lambda_{\mathfrak{q}}:=\tilde{\theta}^{-}Q_{-}$ and recalling that $A_{\omega}:=A_{\omega}^{+-}L_{+-}$, one can exploit the above commutators to compute $W(A_{\omega})=0$ and
\begin{equation}
Z(\star A_{\omega})=\frac{4\tilde{x}^{++}\tilde{x}^{--}}{1+\tilde{y}^2}\biggl[ 1 + \frac{4i\tilde{\theta}^{+}\tilde{\theta}^{-}}{(1-i\tilde{y})[4\tilde{x}^{++}\tilde{x}^{--}+(1+i\tilde{y})^2]}\biggr]\star A_{\omega}^{+-}L_{+-}
\end{equation}
Equation \eqref{EOM-A-WZ} can thus be immediately solved as in \eqref{Solution-EOM} with  $B_{\pm}:= \pm Z^{-1}$ and
\begin{equation}\label{EOM-Solution-OSp}
Z^{-1}=\frac{1+\tilde{y}^2}{4\tilde{x}^{++}\tilde{x}^{--}}\biggl[ 1 - \frac{4i\tilde{\theta}^{+}\tilde{\theta}^{-}}{(1-i\tilde{y})[4\tilde{x}^{++}\tilde{x}^{--}+(1+i\tilde{y})^2]}\biggr]
\end{equation}

\section{Conclusions and Outlook}
We focused on a delicate step in the T-dualisation procedure of symmetric and semi-symmetric cosets, namely solving the EOM for the gauge field $\omega$ in the $\mathfrak{h}$ subspace.
We highlighted how such EOM is not necessarily solvable due to the lack of a linear term in $A_{\omega}$: this may heavily affect T-duality, as the removal of $A_{\omega}$ is necessary to write down the full dual action.
Rewriting the EOM as in \eqref{EOM-A-WZ}, three cases can be recognised in which the solution is of the form \eqref{Solution-EOM}. This has been used to write the full T-dual action and construct solutions for two explicit examples.

An important step toward a better comprehension of T-dualisation would be represented by a deeper understanding of the EOM \eqref{EOM-A-WZ} and of the constraints its resolution imposes on the underlying algebra. Additionally, given the physical relevance of the coset $OSp(1|2)/SO(1,1)$ and the simple result \eqref{EOM-Solution-OSp} for its EOM, it would certainly be very interesting to carry out the dualisation of such model in full details, so as to have a concrete example of super non-Abelian T-duality on supercosets, which could hopefully serve as a base model in view of more complicated ones.  

\begin{acknowledgement}
I am extremely grateful to Silvia Penati, Dmitri Sorokin and Martin Wolf for stimulating discussions and useful comments on this contribution. I also thank the organisers of the conference "2D SUSY@MATRIX" for the opportunity to present my work.
This work was partially supported by Universit{\`a} degli studi di Milano-Bicocca, by the Italian Ministero dell'Universit{\`a} e della Ricerca (MUR), and by the Istituto Nazionale di Fisica Nucleare (INFN) through the research project `Gauge theories, Strings, Supergravity' (GSS).
\end{acknowledgement}

\end{document}